\newcommand{\DTN} {NiCl$_2$$\cdot$4SC(NH$_2$)$_2$\xspace}
\newcommand{\DDTN} {NiCl$_2$$\cdot$4SC(ND$_2$)$_2$\xspace}
\newcommand{\DTNX} {Ni(Cl$_{1-x}$Br$_x$)$_2$$\cdot$4SC(NH$_2$)$_2$ \xspace}

\documentclass[prb,twocolumn,floatfix,preprintnumbers,amsmath,amssymb,superscriptaddress]{revtex4}

\usepackage{graphicx}% Include figure files
\usepackage{dcolumn}% Align table columns on decimal point
\usepackage{bm}% bold math
\usepackage{color}
\usepackage{xspace}

\begin{document}

\title{Critical exponents and intrinsic broadening of the field-induced transition in \DTN}

\author{E. Wulf}
\affiliation{Neutron Scattering and Magnetism, Laboratory for Solid State
Physics, ETH Z\"urich, Z\"urich, Switzerland.}

\author{D. H\"uvonen}
\affiliation{Neutron Scattering and Magnetism, Laboratory for Solid State
Physics, ETH Z\"urich, Z\"urich, Switzerland.}
\affiliation{National Institute of Chemical Physics and Biophysics, Akadeemia tee 23, 12618 Tallinn, Estonia.}

\author{R. Sch\"onemann}
\affiliation{Dresden High Magnetic Field Laboratory (HLD),        Helmholtz-Zentrum Dresden-Rossendorf, D-01328 Dresden, Germany.}

\author{H. K\"uhne}
\affiliation{Dresden High Magnetic Field Laboratory (HLD),        Helmholtz-Zentrum Dresden-Rossendorf, D-01328 Dresden, Germany.}

\author{T. Herrmannsd\"orfer}
\affiliation{Dresden High Magnetic Field Laboratory (HLD),        Helmholtz-Zentrum Dresden-Rossendorf, D-01328 Dresden, Germany.}

\author{I. Glavatskyy}
\affiliation{Helmholtz-Zentrum Berlin f\"ur Materialien und Energie GmbH,
Department Quantum Phenomena in Novel Materials, Berlin, Germany.}

\author{S. Gerischer}
\affiliation{Helmholtz-Zentrum Berlin f\"ur Materialien und Energie GmbH,
Department Quantum Phenomena in Novel Materials, Berlin, Germany.}

\author{K. Kiefer}
\affiliation{Helmholtz-Zentrum Berlin f\"ur Materialien und Energie GmbH,
Department Quantum Phenomena in Novel Materials, Berlin, Germany.}

\author{S. Gvasaliya}
\affiliation{Neutron Scattering and Magnetism, Laboratory for Solid State
Physics, ETH Z\"urich, Z\"urich, Switzerland.}

\author{A. Zheludev}
 \email{zhelud@ethz.ch}
 \homepage{http://www.neutron.ethz.ch/}
\affiliation{Neutron Scattering and Magnetism, Laboratory for Solid State
Physics, ETH Z\"urich, Z\"urich, Switzerland.}

\date{\today}

\begin{abstract}
The field-induced ordering transition in the quantum spin system \DTN is studied by means of neutron diffraction, AC magnetometry and relaxation calorimetry. The interpretation of the data is strongly influenced by a finite distribution of transition fields in the samples, which was present but disregarded in previous studies. Taking this effect into account, we find that the order-parameter critical exponent is inconsistent with the BEC universality class even at temperatures below 100~mK. All results are discussed in comparison with previous measurements and with recent similar studies of disordered \DTNX.

\end{abstract}

 \maketitle

\section{Introduction}

Over the past decade, a great deal of attention has been given to so-called Bose-Einstein condensation of magnons.\cite{Giamarchi2008} These are quantum phase transitions induced in axially symmetric spin systems by the application of an external magnetic field. The simplest example is that of ``conventional'' antiferromagnets at their saturation fields.\cite{Batyev1984} For purely technical reasons, these transitions are easier to study in gapped quantum antiferromagnets, which in zero field have a non-magnetic ground state.\cite{Giamarchi1999} An external field drives the spin gap to zero by virtue of Zeeman effect, at which point spontaneous long-range ordering of transverse spin components may occur, as in the much-studied TlCuCl$_3$ system.\cite{Tanaka2001,Ruegg2003} Quite a few such materials have been studied to date.\footnote{For a recent review see Ref.~\onlinecite{Zapf2014}}

An all too common experimental problem is magnetic anisotropy. In anything but the axially symmetric case, the field-induced transition is of the Ising (rather than BEC) universality class. The result is a re-opening of the gap in the high-field phase\cite{Glazkov2004,Kolezhuk2005,DellAmore2009} and other features not compatible with BEC physics,\cite{DellAmore2009} including unusual values of critical exponents.\cite{Tanaka2001} For this reason, experiments on one particular material, namely \DTN (DTN), have been of special importance.\cite{Paduan2004} This compound is tetragonal, and applying the field along the unique crystallographic axis ensures the required axially-symmetric geometry. A number of studies were aimed at measuring the critical properties of the corresponding BEC quantum critical point. Specifically, studies of the $(H-T)$ phase diagram\cite{Paduan2004,Zapf2006,Yin2008,Weickert2012,Kohama2011} provided data on the so-called crossover exponent $\phi$, which describes the temperature dependence of the critical field $H_c$: $T=[H_c(T)-H_c(0)]^\phi$.\footnote{In literature one often sees the use of the exponent $\alpha=1/\phi$.}  The exponent is expected to have a particular value $\phi=2/3$ for the magnon BEC transition in three dimensions. Several experiments reportedly confirmed this prediction for DTN.\cite{Zapf2006,Yin2008} Other exponents, particularly the order parameter critical index $\beta$ have not been studied experimentally as yet, although the high field ordered state has been investigated with neutron diffraction and inelastic scattering in some detail.\cite{Tsyrulin2013}

DTN gained renewed attention in the context of BEC in the presence of disorder. In experiments on \DTNX  (DTNX), where randomness is introduced on the non-magnetic halogen sites,\cite{Yu2012,Wulf2013} disorder was shown to substantially affect the $(H-T)$ phase diagram. According to Yu {\it et al.},\cite{Yu2012} the crossover exponent changes drastically to $\phi\sim 1$ in the low-temperature regime $T\lesssim 250$~mK.\cite{Yu2012}
This behavior was interpreted in the context of Bose Glass physics,\cite{Fisher1989,Zheludev2013} though there remains a controversy regarding the value of $\phi$ even on the theoretical side.\cite{Yao2014,Fisher1989,Yu2010epl,Yu2012}
Confusingly, recent neutron diffraction experiments have measured the order parameter exponent $\beta$ as well at the crossover exponent $\phi$ in DTNX, but failed to find any indication of Bose Glass behavior.\cite{Wulf2013}

The initial purpose of the present study was to use neutron diffraction and complementary methods to carefully measure $\beta$ and the $(H-T)$ phase boundary in stoichiometric  (disorder-free) \DTN, for a direct comparison with previous results on  DTNX.\cite{Wulf2013} Our main finding is  that our measurements, as well as all previous studies, are significantly influenced by a {\it distribution of transition fields} in the sample. Taking this effect into account, for DTN we find a critical exponent $\beta$ that is {\it not consistent with the BEC universality class} all the way down to  $T<100$~mK. Moreover, we come to the conclusion that the previously reported\cite{Yin2008} $\phi\sim 2/3$ is likely due to an inappropriately wide choice of fitting range, while the actual data support a much larger value for $T<170$~mK. Finally, comparing the results on DTN and DTNX, we conclude that there is no statistically significant evidence of any effect of disorder on the critical properties at low temperatures. These findings cast doubt on the relevance of the BEC and Bose Glass paradigms to the realities of DTN and DTNX.

\section{Experimental}
Single crystal samples of DTN for the present study were grown from solution using the thermal gradient method, as  in Refs.~\onlinecite{Paduan2004,Zapf2006,Yin2008,Weickert2012,Kohama2011}. It is important to stress that virtually all previous studies used samples from the same source,\cite{Paduan2004} but the crystals for this work were grown independently. All DTN material used in the present study was fully deuterated, to facilitate neutron diffraction measurements. The crystal structure was verified using single crystal X-ray diffraction on a Bruker AXS diffractometer equipped with an  APEX-II detector, and found to be indistinguishable from that of protonated DTN to within accuracy of the instrument.

\begin{figure}
\unitlength1cm
\includegraphics[width=\columnwidth]{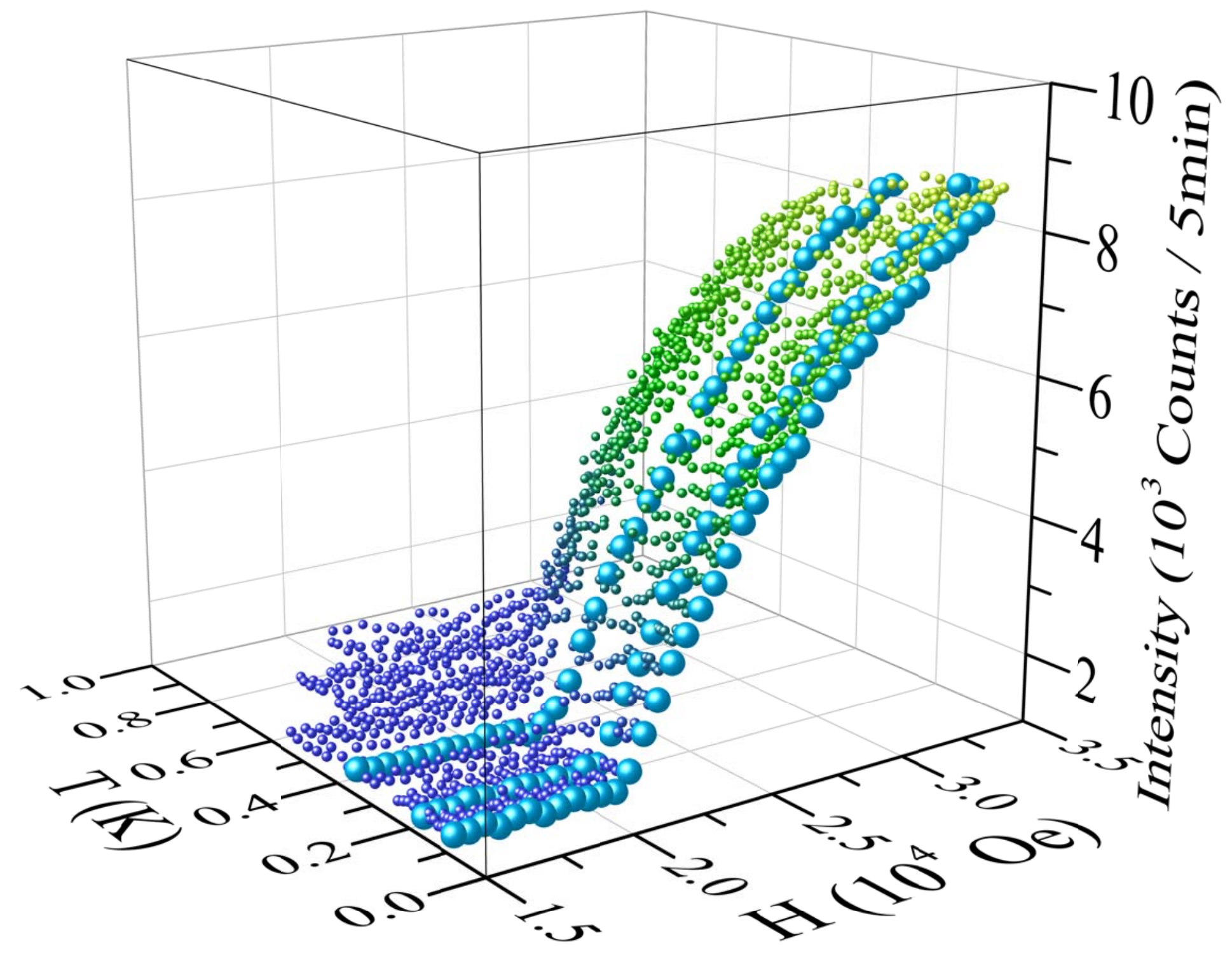}
\caption{The bulk of neutron diffraction peak intensity data measured in constant-temperature field sweeps in \DDTN at the $(0.5,0.5,0.5)$ reciprocal space point.  Larger symbols highlight the scans shown in more detail in Fig.~\ref{neutron}.\label{alldata}}
\end{figure}

Neutron diffraction was performed at the E2 diffractometer at Helmholz-Zentrum Berlin with $\lambda= 2.38$~\AA~ neutrons. The single-crystal sample had the size $7\times 7\times 6$~mm$^3$ and mosaic spread of 1$^\circ$~full width at half maximum (FWHM). We used a $^3$He-$^4$He dilution refrigerator in a 4.2~T superconducting split-coil magnet. The crystal alignment was verified in-situ. The measured angle between the field direction and the crystallographic $c$ axis was 1.7~$^\circ$.  Most data were collected in the vicinity of the $(0.5,0.5,0.5)$ reciprocal-space point. It corresponds to the smallest-angle magnetic Bragg peak in the high-field ordered phase.\cite{Tsyrulin2013} The measured Bragg widths were in all cases resolution-limited. The corresponding peak intensity was collected by sweeping the magnetic field at constant temperature. The bulk of the data are visualized in Fig.~\ref{alldata}.

Thermal-relaxation calorimetry was performed on a Quantum Design PPMS with the $^3$He-$^4$He dilution refrigerator insert. A deuterated single crystal sample with a mass of 1.7~mg was aligned using X-ray diffraction. The misalignment of the applied field with the crystallographic $c$ axis was smaller than 5$^\circ$.

The complex magnetic AC susceptibility of \DDTN was measured  by using a compensated mutual inductance mount to the mixing chamber of a $^3$He-$^4$He dilution refrigerator placed in the bore of a superconducting magnet system. The sample was a 52~mg single crystal attached to a silver holder by a small amount of vacuum grease and aligned to better that 5$^\circ$ relative to the field direction.  The temperature- and field-dependent AC susceptibility was recorded by applying a primary AC field with $\mu$T amplitude at a frequency of 2~kHz to the sample. For that, we used a Stanford Research SR 830 lock-in amplifier to sense the induced picked-up voltage in the compensated secondary coil pair.
The AC field was superimposed by axially co-aligned static fields up to 14~T.

\section{Results and initial data analysis}

\begin{figure}
\unitlength1cm
\includegraphics[width=\columnwidth]{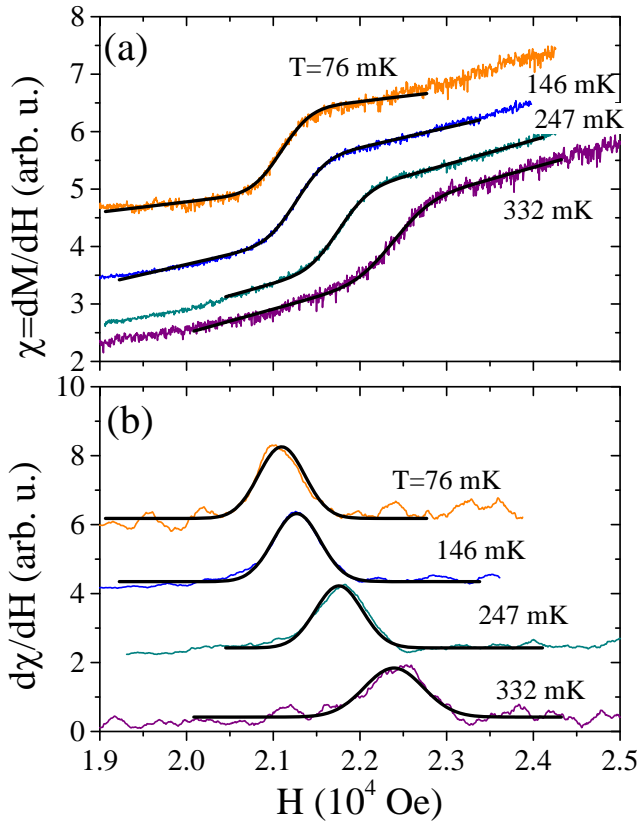}
\caption{Thin lines: typical field dependence of magnetic susceptibility $\chi=dM/dH$ (a) and its field derivative $d\chi/dH$ (b) measured in \DDTN at several temperatures. The thick lines are Gaussians in (b) which corresponds to an error function on a sloping background in (a). Except for the highest temperature, all data are plotted with arbitrary offsets for visibility.\label{chi}}
\end{figure}

\subsection{Magnetic susceptibility}
Typical constant temperature measurements of magnetic AC susceptibility are shown in Fig.~\ref{chi}a (thin lines). The transition is marked by a distinct step in $\chi(H)$, in agreement with previous studies.\cite{Paduan2004} The step is visibly broadened at all temperatures. Its shape can be approximated by the error function. The field derivative  of the measured susceptibility $d \chi/dH$ is plotted in Fig.~\ref{chi}b (thin lines), for a direct comparison with Fig.~2a from Ref.~\onlinecite{Yin2008}, where the raw data look very similar.

\begin{figure}
\unitlength1cm
\includegraphics[width=\columnwidth]{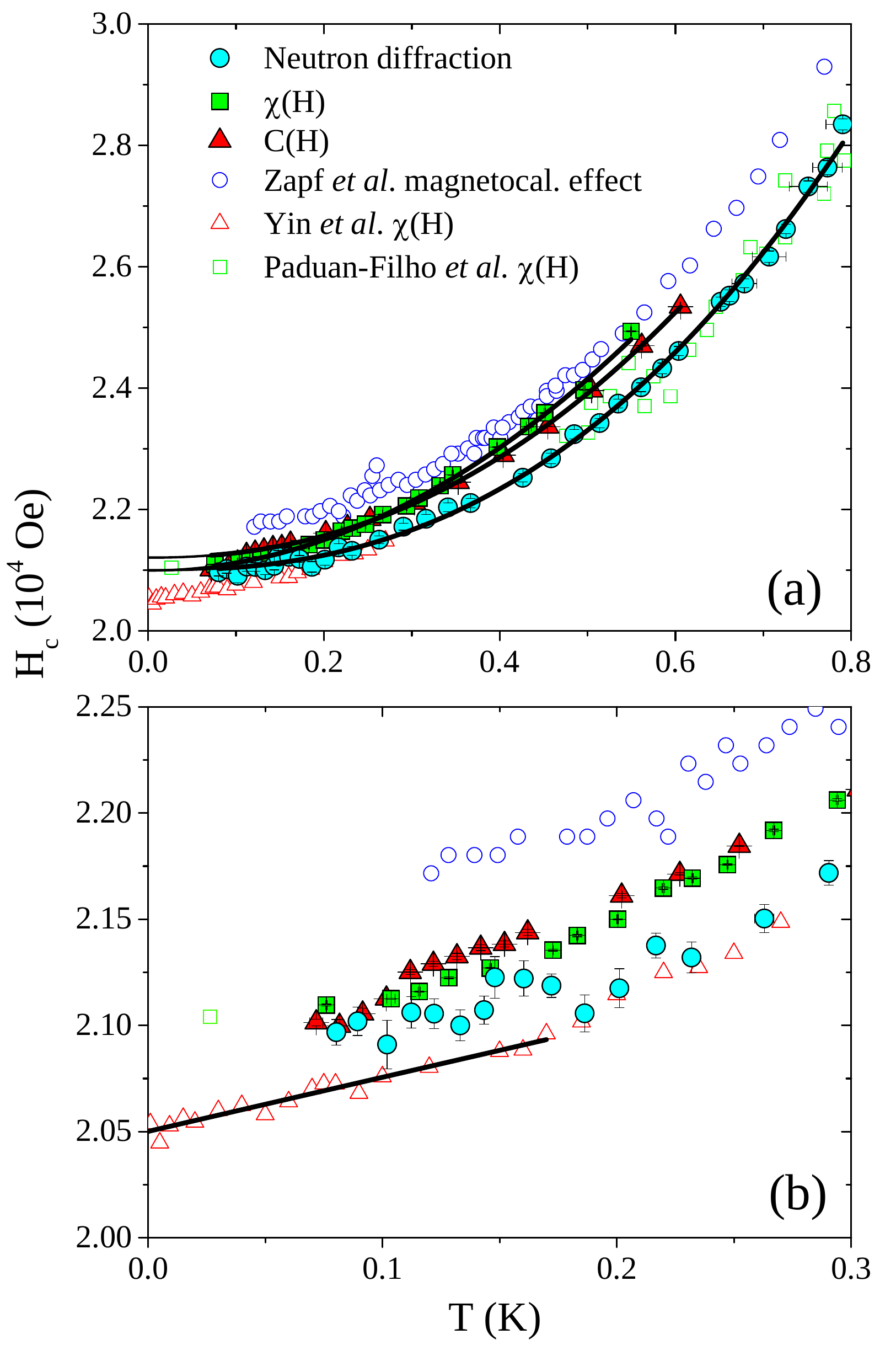}
\caption{(a): Magnetic field-temperature phase diagram of DTN. Circles (neutron diffraction), squares (magnetic AC susceptibility) and triangles (calorimetry) are data obtained in this work on deuterated samples. Open squares,\protect\cite{Paduan2004} circles\protect\cite{Zapf2006} and triangles\protect\cite{Yin2008} are results for protonated DTN reported in literature. (b): The low temperature data shown in more detail. The solid lines in (a) are power law fits in a wide temperature range. The solid line in (b) is a power law fit to the data of Ref~\onlinecite{Yin2008} up to $T_\mathrm{max}=170$~mK, yielding $\phi=1.00(14)$.\label{phase}}
\end{figure}

\begin{figure}
\unitlength1cm
\includegraphics[width=\columnwidth]{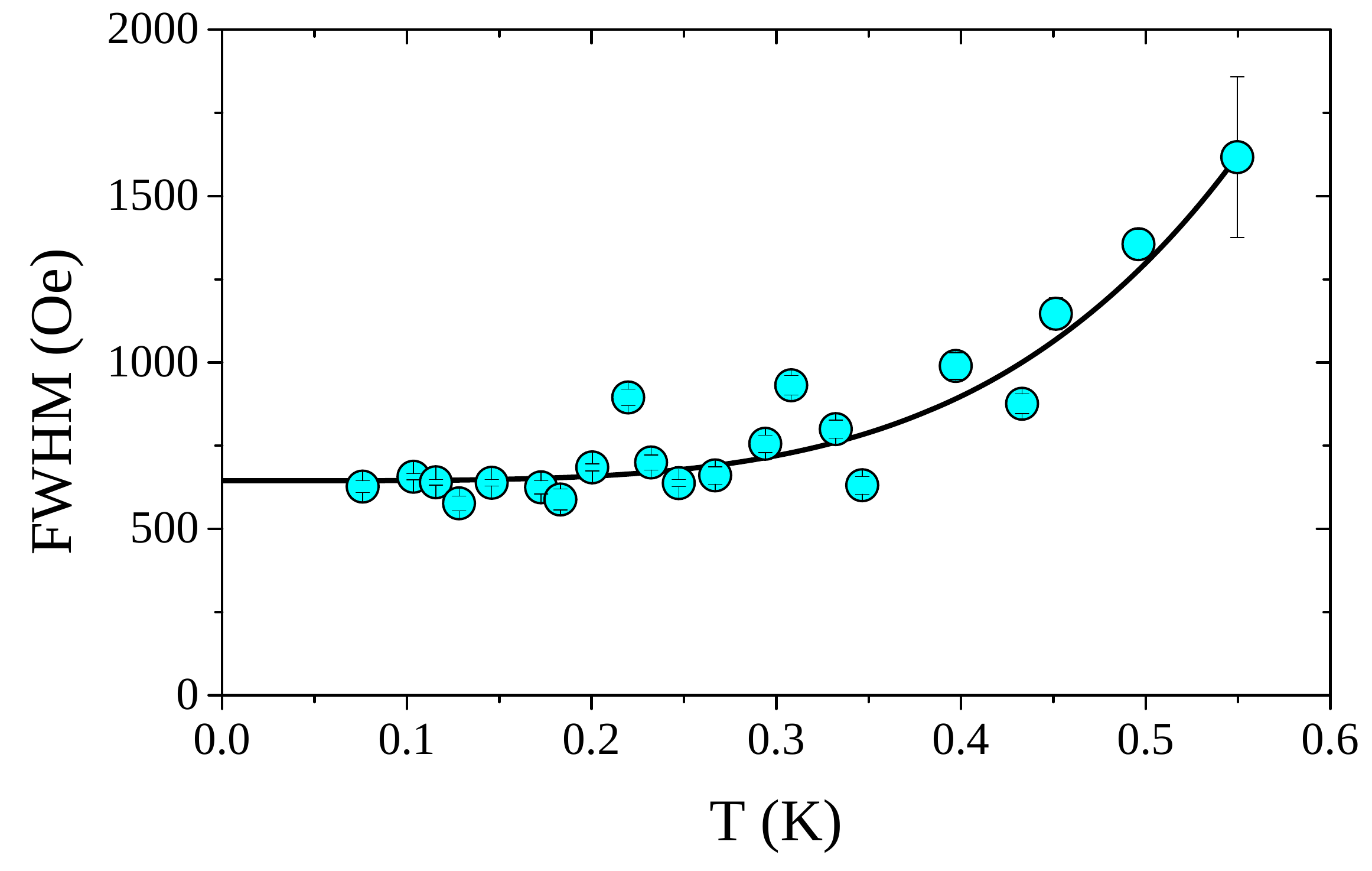}
\caption{Symbols: temperature dependence of the peak in $d\chi/d_H$ in \DDTN, as determined by Gaussian fits described in the text. The solid line is a guide for the eye only.\label{width}}
\end{figure}

Our $\chi(H)$ data were analyzed using fits with an error function on a linear sloping background (Fig.~\ref{chi}a, heavy solid curves), which corresponds to Gaussian in $d\chi(H)/dH$ (Fig.~\ref{chi}b, heavy solid curves). The center of the Gaussian determined at each temperature is shown in solid squares in Fig.~\ref{phase}. The FWHM of peak $\Delta H$ is plotted versus temperature in  Fig.~\ref{width}. $\Delta H$ remains constant  below $\sim 300$~mK, and gradually increases at higher temperatures.

The observed transition width  exceeds any expected field inhomogeneities or inaccuracies in setting the field value, all estimated to be below 5~mT.
The broadening is also not a finite-$T$ effect, as its temperature dependence totally levels off at $T\rightarrow 0$. We conclude that this width is an {\it intrinsic} feature of the samples studied.  In fact, in our case it is about 30\% {\it narrower} than in the experiments of Refs.~\onlinecite{Yin2008} and \onlinecite{Paduan2004}.

As will be discussed in detail below, we interpret the apparent width of the transition as a {\it distribution of transition fields $H_c$ in the sample}.  In the analysis of the neutron and calorimetry data, we shall approximate this distribution as normal (Gaussian), with FWHM equal to that of $d \chi/dH$ at base temperature: $\Delta H_c= 650$~Oe.

\subsection{Neutron diffraction}
\begin{figure}
\unitlength1cm
\includegraphics[width=\columnwidth]{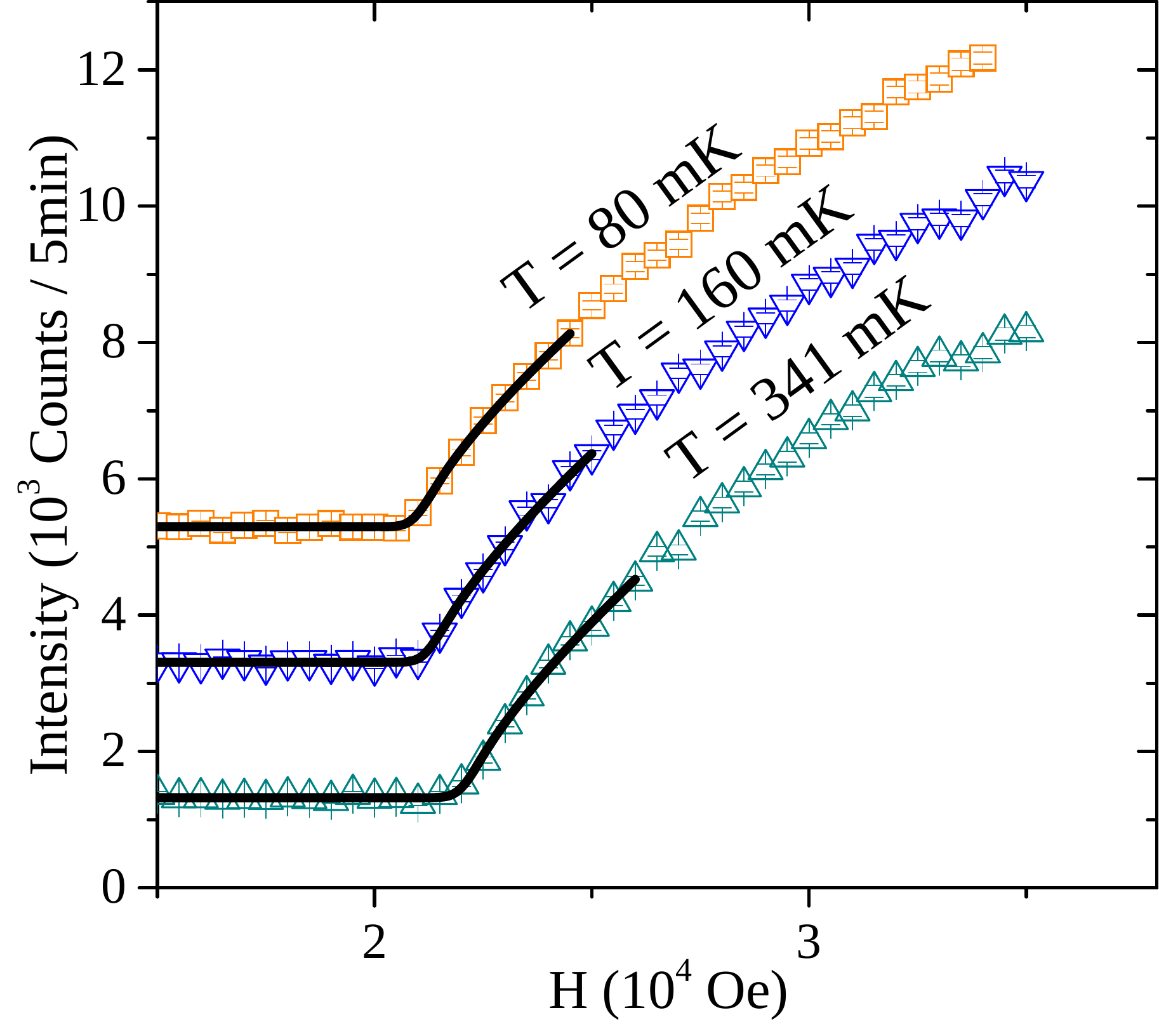}
\caption{Symbols: typical field dependencies of the neutron diffraction peak intensity at the $(0.5,0.5,0.5)$ reciprocal-space position measured in \DDTN at several temperatures. Solid lines are fits to a power law function convoluted with a Gaussian distribution of transition fields, as described in the text. In all cases the fitting range is 4~kOe above $H_c$. All data for 160~mK and 80~mK are plotted with offsets of $2\cdot 10^3$~counts/5~min and  $4\cdot 10^3$~counts/5~min, respectively. \label{neutron}}
\end{figure}

Some representative field-sweeps of the $(0.5,0.5,0.5)$ magnetic Bragg peak intensity are shown in Fig.~\ref{neutron}. Similarly to what was previously done for DTNX,\cite{Wulf2013} and following the procedure outlined in Ref.~\onlinecite{Huevonen2012}, each such data set was analyzed using power-law fits in a progressively shrinking field window. The fit parameters were the critical feld $H_c$, the exponent $\beta$, and an overall scale factor. The difference in our present approach was that instead of using a ``bare'' power law function, we convoluted it with a Gaussian distribution of transition fields with a fixed  FWHM $\Delta H_c=650$~Oe, as discussed above.  The convoluted function gives comparable fits to the data and reproduces the slight ``rounding'' of the transition clearly visible in some field scans.

Following Ref.~\onlinecite{Huevonen2012}, for each field sweep, we identified the ``narrowest useful fitting range'' around the transition point, defined by the maximum field $H_\mathrm{max}$ used in the fits. Decreasing the fitting range further does not lead to a statistically significant change in the fitted parameter values, while the error bar increases. Having compared $H_\mathrm{max}-H_c$ for all temperatures studied, we chose a common fitting range for all data sets: $H_\mathrm{max}=H_c+\delta$, $\delta=4$~kOe.

\begin{figure}
\unitlength1cm
\includegraphics[width=\columnwidth]{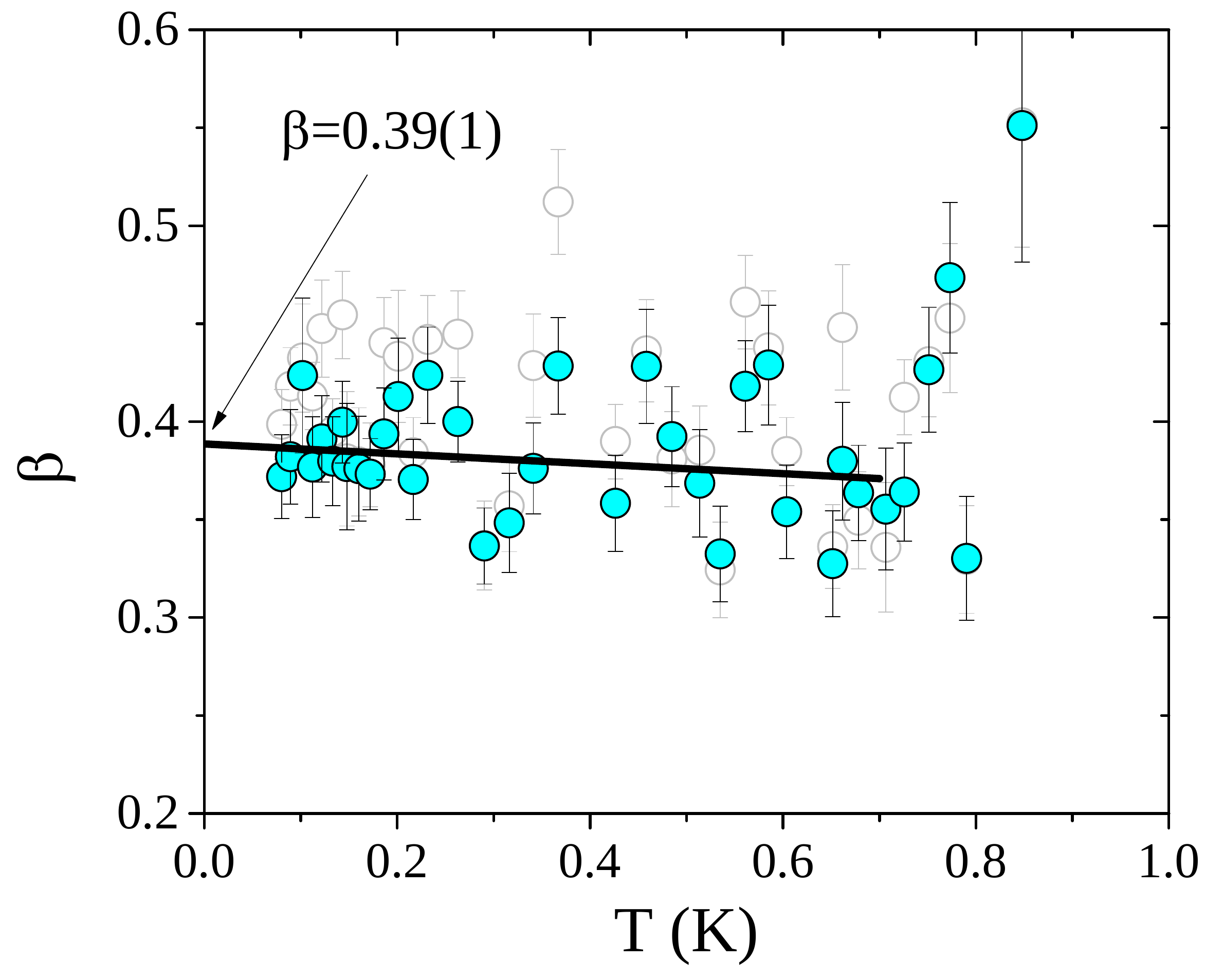}
\caption{Exponent of the power law describing the field dependence of the order parameter in \DDTN, plotted as a function of temperature. Solid symbols represent fits that take the finite distribution of critical fields into account. For comparison, open symbols are fits using a ``bare'' power law function. In all cases the fitting range in field in 0.4~T above the transition. The line is a linear fit to the solid symbols below $700$~mK.\label{beta}}
\end{figure}

Analyzing all field sweeps in the common fit range gives the temperature dependence of $H_c$ and $\beta$ shown in solid circles in  Figs.~\ref{phase} and \ref{beta}, respectively. For comparison, in Fig.~\ref{beta} open circles show the values obtained using ``bare'' power law fits in the same range.

\subsection{Calorimetry}
\begin{figure}[!htb]
\unitlength1cm
\includegraphics[width=\columnwidth]{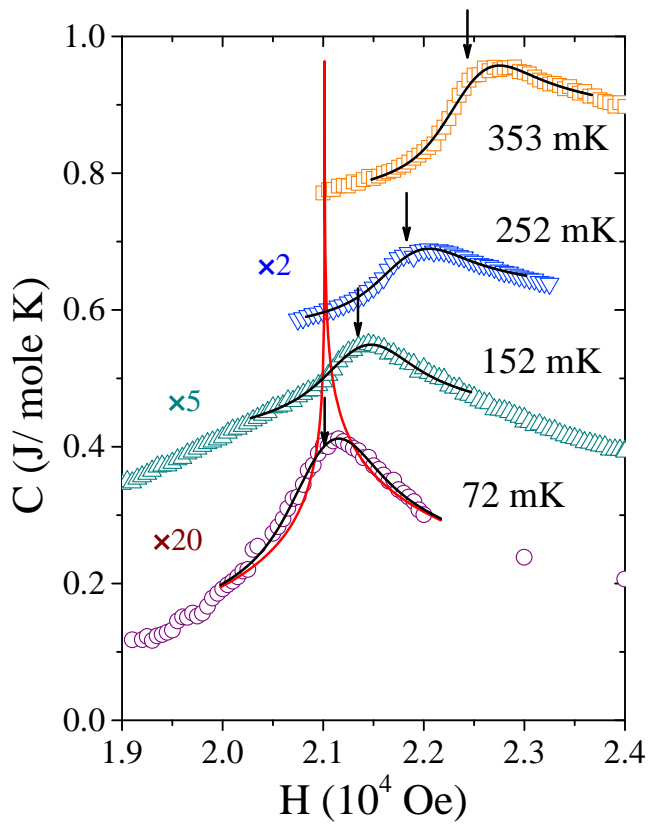}
\caption{Symbols: typical field dependencies of specific heat measured in \DDTN at several temperatures. The solid lines through the data points are fits to the data using a power law convoluted with a finite distribution of transition fields, as described in the text. The sharply peaked curve is the``bare'' non-convoluted power law for the $T=72$~mK data set. Arrows are the fitted transition fields.\label{C}}
\end{figure}

Typical measured field dependencies of specific heat are shown for different temperatures in Fig.~\ref{C}. The transition is marked by a well-defined maximum. Compared to similar features in other organic gapped quantum magnets at a field-induced ordering transition,\footnote{See, for example, typical data for H$_8$C$_4$SO$_2\cdot$Cu$_2$Cl$_4$ in Ref.~\onlinecite{Wulf2011}} the peak is somewhat broadened. Following the logic of the discussion above, we attributed this broadening to a finite distribution of transition fields, and analyzed the data accordingly. Our model for each field scan at a constant temperature was based on a power law function, with the critical exponent $\alpha=-0.015$ for a thermodynamic XY transition in three dimensions.\cite{Pelissetto2002} For each temperature, the parameters of the fit were $H_c$, two scales factors for $C(H)$ below and above the transition, respectively, and an overall flat background.

The power law function was numerically convoluted with a normal distribution of fixed FWHM $\Delta H_c=650$~Oe. In all cases we used a fitting range of $+1250/-850$~Oe around the peak position. Good fits are obtained at all temperatures. Typical fits are shown as smooth curves in Fig.~\ref{C}. As a reference, the sharply peaked curve is the ``bare'' (non-convoluted) power law function corresponding to the fit curve for $T=72$~mK. We see that even if there is a lambda anomaly at the transition, it is totally masked by the critical field distribution.

The arrows in Fig.~\ref{C} indicate the fitted value of $H_c$ at each temperature. Note that due to the convolution effect, it is always below the apparent specific heat maximum. The temperature dependence of the critical field obtained in our analysis is plotted in solid triangles in Fig.~\ref{phase}.

\section{Discussion}

\subsection{Phase boundaries and transition width}
What is immediately noticeable are significant differences between phase boundaries measured with different techniques. The discrepancies between our measurements and previous studies of the protonated material may be at least in part due to us using deuterated DTN in all experiments. However, this circumstance can not account for the differences in phase boundaries that we measure with neutrons, magnetometry and calorimetry in deuterated DTN, or between previous magnetometry\cite{Yin2008,Paduan2004} and magnetocaloric\cite{Zapf2006} studies of protonated crystals. A point of concern is the orientation of the magnetic field in the sample. The phase diagram in fields applied in the $(a,c)$ plane, at an angle to the unique tetragonal $c$ axis, has been thoroughly investigated by neutron diffraction.\cite{Tsyrulin2013} Approximating that measured phase boundary by an ellipse, we conclude that even a 5$^\circ$  misalignment will change the critical field by only about 100~Oe. Therefore, sample alignment is not an issue in any of our or previous studies.

As pointed out in Ref.~\onlinecite{Weickert2012}, a much larger concern is mechanical stress in the sample. Those measurements have shown that the transition field in DTN is exceptionally sensitive to pressure. Indeed, it was estimated that the stress produced on the sample by a rather gentle dilatometer spring may alter $H_c$ by as much as 300~Oe. The stress may be much larger for the samples used in our calorimetry experiments and magnetometry. They are attached to the calorimeter platform with vacuum grease that is bound to stress the sample upon cooling due to a different thermal expansion coefficients. The lowest stress occurs in our neutron experiments, where a large sample is mounted without any glue, but is simply held in place with thin Al wire.

Stress, specifically intrinsic residual stress due to defects, is also the most likely explanation for the observed {\it broadening of the transition}. Since all DTN samples are solution grown, they all inevitably have cracks, imperfections and solvent inclusions.\cite{crystalhandbook} A microscope inspection of our samples reveled numerous defects of this type.\footnote{All samples used in Ref.~\onlinecite{Wulf2013} have similar defects.} Upon cooling, such defects, particularly solvent inclusions subject to freezing, will most certainly generate a distribution of strong strain fields in the sample, resulting in a distribution of critical fields. The broadening will be sample-dependent. As mentioned above, in our samples it is about 30\% {\it smaller} than reported in previous studies.

\subsection{Critical exponent $\beta$}
One of the main results of this work are measurements of the magnetic order parameter. Compared to previous studies,\cite{Tsyrulin2013} assuming that the transition is continuous and described by a power law, our data provide enough statistics to extract the corresponding critical index $\beta$.  As shown in Fig.~\ref{beta}, below $\sim 0.7$~K the experimental value slowly increases with decreasing temperature. Averaged over the upper 100~mK of this range, $\beta=0.36(1)$, which is fully consistent with the expectations for a thermodynamic XY transition in three dimensions.\cite{Pelissetto2002} The temperature dependence could be interpreted as a crossover to the quantum critical regime at low temperatures. However, linearly extrapolating the measured exponent to $T=0$ (see solid line in Fig.~\ref{beta}), we get $\beta=0.39(0.01)$. This value is {\it inconsistent with the mean field expectation for 3-dimensional BEC of magnons, $\beta=0.5$}. If there is a crossover to larger values, it occurs at still lower temperatures.

How reliable is this conclusion? As always in a diffraction experiment, the main potential pitfall is that near the transition some critical fluctuations are picked up due to the finite resolution of the instrument. However, this has always the effect of increasing the  intensity near $H_c$, resulting in {\it larger} apparent critical indexes, and can not explain the discrepancy. Another concern is whether the field range has been chosen appropriately, {\it i.e.},  sufficiently narrow to access the quantum critical regime. This indeed was a problem for the previous studies of DTNX,\cite{Wulf2013} where even at $T\rightarrow 0$ we expected a crossover {\it vs. field} from BEC behavior to that dominated by disorder. In disorder free DTN, the only expected crossover is to the classical (thermodynamic) regime at $T>0$. Fortunately, the field range of the classical transition rapidly tends to zero as $T^{1/\phi}$, with the same power law exponent as the phase boundary.\cite{Sachdevbook} With the assumption that the field-width of the classical region is of the same order as the change in $H_c$ compared to zero temperature, for DTN at 200~mK it is narrower than 500~Oe. Our fits over a range of 4~kOe are therefore not affected. At the same time, the maximum Bragg intensity used in our fits is still three times smaller than the saturation value.\cite{Tsyrulin2013} Thus, saturation effects are also unlikely to influence the analysis. We conclude that {\it assuming the transition is continuous and described by a power law}, our estimate of $\beta$ is quite robust.

\subsection{Critical exponent $\phi$}
Previous studies of the phase boundary in DTN in temperatures up to 1~K, have given a critical exponent $\phi\sim 0.4$.\cite{Paduan2004} This result is generally consistent with power-law fits to our neutron diffraction data up to 0.8~K ($\phi=0.41(1)$, $H_c=21.0(1)$~kOe), calorimetry up to 0.6~K ($\phi=0.45(1)$, $H_c=21.2(1)$~kOe) and susceptibility  up to 0.55~K ($\phi= 0.50(1)$, $H_c=21.0(1)$~kOe). The corresponding fits are shown in solid lines in Fig.~\ref{phase}a. While these values appear to be at odds with the prediction $\phi=2/3$ for a 3-dimensional BEC transition, a likely reason for the discrepancy was discussed in Ref.~\onlinecite{Yin2008}. It was suggested that the BEC value for the crossover exponent is recovered only for the lowest temperatures, below $T\sim 270$~mK. At higher temperatures one observes classical (thermodynamic) critical behavior which, with a corresponding change in $\phi$.

All our experiments lack sufficient data at the lowest temperatures for a reliable power-law analysis in this regime. An additional problem is posed by the distribution of critical fields. A  650~Oe variation of $H_c$ in the sample corresponds to a $\sim 200$~mK variation of $T_c$. Any measurements of the ``critical exponent'' $\phi$ in this range may be strongly affected. In fact, the transition field at each temperature can not be unambiguously extracted from the data without an implicit assumption regarding the {\it shape} of the singularity in the susceptibility. For example, taking the maximum of the derivative $d\chi/dH$ as a measure of the ``average'' $H_c$, as was done here and in  Ref.~\onlinecite{Yin2008}, {\it implies} a BEC-like step of $\chi(H)$ and a symmetric distribution of transition fields. A rather {\it different} definition of $H_c$ was used in Ref.~\onlinecite{Paduan2004}, and may potentially lead to a different extraction of $\phi$.
The ambiguity becomes acute if the shape of the measured $\chi(H)$ curve is itself temperature-dependent. Fortunately, for DTN, at low temperatures this does not appear to be the case. However, for DTNX, $\chi(H)$ curves become visibly broadened at low temperatures,\cite{Yu2012} which may severely impact the determination of $\phi$.

For DTN, at best, one can {\it assume} the transition corresponds to BEC, analyze the data accordingly, and check whether the outcome is consistent with the BEC interpretation. In this spirit, analyzing the data measured up to $T_\mathrm{max}=270$~mK , the authors of Ref.~\onlinecite{Yin2008} obtained $\phi=0.68(1)$, in excellent agreement with BEC. However, given that at higher temperatures there is a crossover to $\phi\lesssim0.5$, is 270~mK is low enough? Apparently not. For the {\it same} data, reducing $T_\mathrm{max}$ to 170~mK gives a {\it statistically significant} change in the fitted value: $\phi=1.00(0.14)$ (solid line in Fig.~\ref{phase}b).\footnote{ Further reducing $T_\mathrm{max}$ does not lead to an appreciable change in the fitted parameter values.} This behavior is consistent with the obvious observation that the data in Fig.~3b of Ref.~\onlinecite{Yin2008} appear {\it almost linear} at low temperatures. The same is actually true for the upper critical field $H_{c2}$ as well, as shown in Fig.~3a of Ref.~\onlinecite{Yin2008}. In fact, for $H_{c2}$ the authors note an ``abnormal change in slope'' at 150~mK, but fail to note a very similar feature in the lower critical field. In this context, for both critical fields, the data of Ref.~\onlinecite{Yin2008} point to $\phi \sim 1$ at low temperatures, with a crossover to $\phi\sim 0.4$ at high temperatures. The reported $\phi\sim 2/3$ simply corresponds to an {\it accidental choice of fitting range} and a crossover between the two regimes.

\subsection{Comparison with DTNX}
Our neutron diffraction data for DTN can be directly compared to those for DTNX (8\% Br) reported previously.\cite{Wulf2013} Apart from the substantial difference in critical field ($H_c\sim 12$~kOe in DTNX vs. $H_c\sim 21$~kOe in DTN), the behavior observed in the two compounds is remarkably similar.  In DTNX, the observed exponent $\beta$ is somewhat larger, $\beta=0.52(3)$,\cite{Wulf2013} vs. DTN's $\beta=0.39(1)$. However, this discrepancy may be due to limitations in the analysis of DTNX data: i) a much larger fitting range $\delta=10$~kOe and ii) not taking into account the finite distribution of critical fields. Indeed, by roughly estimating the width of the transition from the measured specific heat curves in DTNX, and then using the result to re-analyze the corresponding neutron  data, for  8\% DTNX at $T<300$~mK we obtain $\beta= 0.32(5)$.\footnote{ Unfortunately, for DTNX it is not possible to separate the finite distribution of critical fields due to a macroscopic strain inhomogeneity from the intrinsic microscopic effects of Br-Cl disorder. Thus, $\beta= 0.32(5)$ should be considered a lower bound on the actual exponent. The value $\beta=0.52(3)$ obtained without regard for macroscopic inhomogeneity is a very conservative upper bound. }

At this point, the {\it only} experimentally observed effect of  disorder on the field-induced phase transition in DTN is the claim of Ref.~\onlinecite{Yu2012} that {\it in the low-temperature limit} $\phi$ changes from $\phi\sim 2/3$ in the pure material (as estimated in Ref.~\cite{Yin2008}) to $\phi \sim 1$ in DTNX. As discussed above, the former result is an artifact of an accidentally selected fitting range. In fact, the existing data suggest $\phi\sim 1$ in disorder-free DTN {\it as well}. Of course, this value of $\phi$ seems to govern the phase boundary in DTN below $\sim 300$~mK, and only below $\sim 170$~mK in pure DTN. This difference in range is, however, not unexpected. These are different materials, after all, with even critical fields differing by a factor of two.

Thus, there is {\it no statistically significant evidence} of that disorder is at all relevant for the phase transition in DTNX. The only real difference is the value of $H_c$. It is most likely due to a change in the exchange parameters and anisotropy, due to the effect of ``chemical pressure''.  Considering the extreme sensitivity of $H_c$ to strain,\cite{Zapf2011,Weickert2012} a substantial change of $H_c$ on Br-``doping'' is only to be expected.

\subsection{Is the transition continuous?}

It is often overlooked that, based on very robust symmetry arguments, the field-induced transition in DTN is necessarily {\it not} a BEC of magnons, but a {\it discontinuous} transition.\cite{DellAmore2009} In a tetragonal crystal, due to magnetoelastic coupling, a 1st-order transition that lifts the crystal symmetry occurs just before the spin gap closes. The spin gap never closes completely, and increases again beyond the transition, since broken tetragonal symmetry implies Ising-like anisotropy in the system. This mechanism for a quantum phase transition is akin to the famous argument of Larkin and Pikin that certain magnetic thermodynamic transitions involving a coupling to acoustic phonons must be discontinuous.\cite{Larkin1969} Usually, the hope is that the discontinuity is very slight, in which case the transition may be studied as a continuous one. However, for DTN, where magnetoelastic coupling is enormous,\cite{Zapf2011} there are no obvious theoretical grounds for such optimism.

Due to the ever-present distribution of transition fields, the question of the continuity of the transition appears impossible to resolve experimentally. Simply put, a continuous distribution of transition fields will lead even a discontinuous transition to look continuous. Magnetic-susceptibility measurements are hardly an indicator since $\chi(H)$ is discontinuous already in ideal BEC case for the $T\rightarrow 0$ limit. Specific heat may have been more sensitive to the continuity of the transition. However, the weak lambda anomaly in the case of small negative $\alpha$, when convoluted with the normal distribution of transition fields, is indistinguishable from a discontinuous step function. A further complication is that such a discontinuity will be superimposed on a peak-like feature due to an {\it almost} vanishing spin gap in the vicinity of $H_c$.

Neutron diffraction could in principle  provide the most direct measure of the discontinuity. However, a finite $H_c$ distribution will smear out any jump of the Bragg peak intensity as well. From the known $\Delta H_c$, we can say with certainty that a 5\% jump of Bragg peak intensity compared to the saturation value\cite{Tsyrulin2013} would be undetectable in our data. Since Bragg intensity scales as the square of the ordered moment, this corresponds to an order parameter jump {\it as large as 20\% of saturation}! The same inhomogeneity of residual stress will ensure an abundance of nucleation sites and eliminate any hysteresis even for a discontinuous transition.

As a side remark, discontinuous magnetoelastic transition of the type described in Ref.~\onlinecite{DellAmore2009} would be the most natural explanation to another established feature of DTN, namely the {\it gap} in the spin excitation spectrum observed with both ESR\cite{Zvyagin2008} and neutron spectroscopy.\cite{Tsyrulin2013} To date, it is not clear whether there are Goldstone modes in addition to the gap mode as follows from the model in Ref.~\onlinecite{Zvyagin2008}, or whether the system is truly gapped. The latter scenario is exactly what follows from symmetry arguments, but would be totally inconsistent with BEC.

\section{Conclusion}
The above discussion can be summarized as follows:
\begin{enumerate}
\item A finite distribution of transition fields  is a major complication for any experimental studies of DTN. Without specific assumptions regarding the functional form of singularities at the transition, a meaningful discussion of critical exponents in DTN in the low-temperature regime is hardly possible. In fact, none of the new or previously published data can even prove that the transition is continuous. The order parameter jump may be absent, but may also be as large as 20\% of saturation.
\item {\it Assuming} a continuous field-induced transition at temperatures below $\sim 170$~mK, our neutron diffraction data for the order parameter exponent $\beta$, as well as the previously published data for the phase boundary,\cite{Yin2008} are {\it not consistent} with the BEC universality class.
\item Neither the neutron experiments on DTNX and DTN, nor the previously published susceptibility data contain {\it any evidence} of that disorder significantly influences the critical behavior below $T\sim 170$~mK. In particular, the assertion that BEC criticality in DTN gives way to Bose Glass criticality in DTNX, at present lacks experimental justification.
\end{enumerate}

All in all, DTN demonstrates some fascinating physics at low temperatures, but appears not to be a particularly good BEC prototype system, by any measure.

\acknowledgements
This work was partially supported by the Swiss National Science Foundation,
Division 2, and by the European Commission under the 7th framework program through the ``Research Infrastructures'' action of the ``Capacities'' program, CP-CSA-INFRA-2011-1.1.17 Number 233883 NMI3 II. Additional support was provided by  the Estonian Ministry of Education and Research under grant IUT23-03 and Estonian Research Council grant PUT451, as well as by the
Deutsche Forschungsgemeinschaft (DFG) through the Research Training Group GRK 1621. We thank V. Zapf for providing the data plotted in Fig.~3 of Ref.~\onlinecite{Yin2008}

\bibliography{../../../../bib/azbib}

\end{document}